\documentclass[12pt]{article}
\usepackage{latexsym,graphicx,multirow}
\usepackage{float}
\usepackage{amssymb}
\usepackage{amsmath}
\usepackage{amscd}
\usepackage{amsthm}
\usepackage[left=2cm,top=2.5cm,right=2.5cm,bottom=1.5cm]{geometry}
\usepackage{hyperref}
\usepackage{epstopdf}
\usepackage{cite}
\usepackage{caption}
\usepackage{subcaption}
\usepackage{color}
\usepackage[utf8]{inputenc}
\usepackage[T1]{fontenc}
\begin{document}
	
\begin{center}
\large{\bf{ Corrected holographic dark energy with power-law entropy and Hubble Horizon cut-off in FRW Universe  }} \\
		\vspace{10mm}
\normalsize{ Vinod Kumar Bhardwaj$^1$, Priyanka Garg$^2$, Anirudh Pradhan$^3$, Syamala Krishnannair$^{4}$}\\
		\vspace{5mm}
\normalsize{$^{1,2}$Department of Mathematics, Institute of Applied Sciences \& Humanities, GLA University,\\ 
	Mathura -281 406, Uttar Pradesh, India} \\
\normalsize{$^{3}$Centre for Cosmology, Astrophysics and Space Science (CCASS), GLA University,\\ 
	Mathura -281 406, Uttar Pradesh, India} \\
	\normalsize{$^{4}$Department of Mathematical Sciences, Faculty of Science, Agriculture and Engineering,University of 
		Zululand, Kwadlangezwa 3886, South Africa} \\
	\vspace{2mm}
	$^1$E-mail: dr.vinodbhardwaj@gmail.com \\
	\vspace{2mm}
	$^2$E-mail:pri.19aug@gmail.com \\
	\vspace{2mm}
    $^3$E-mail:pradhan.anirudh@gmail.com  \\
    \vspace{2mm}
              $^{4}$E-mail:krishnannairs@unizulu.ac.za\\
	\vspace{10mm}
		
\end{center}
	
\begin{abstract}
	In the present work, we investigate the power-law entropy corrected holographic dark energy (PLECHDE) model with Hubble horizon
cutoff. We use 46 observational Hubble data points in the redshift range $0 \leq z \leq 2.36$ to
determine the present Hubble constant $H_0$ and the model parameter $n$. It represents a phase transition of the universe from deceleration to 
acceleration and has the transition point at $z_t = 0.71165$. We investigate the observational constraints on the model and calculate some 
relevant cosmological parameters. We examine the model's validity by drawing state-finder parameters that yield the result compatible with the modern
observational data. The model's physical and geometrical characteristics are also explored, and they are shown to match well with current
observations of observational Hubble data (OHD) and the latest joint light curves(JLA) datasets.

\end{abstract}
	
	

\section{Introduction}
We begin with the famous quote by Allan Sandage \cite{ref1} that All the observational cosmological models are in search of 
two parameters: Hubble $H_0$ and deceleration parameters $q_0$. The universe is a dynamic system in which its constituents (galaxies) 
travel like a disciplined march of soldiers and move away from each other with Hubble's rate. The recent astrophysical measurements of 
OHD (Observational Hubble data), SN Ia (Type Ia supernovae) \cite{ref2,ref3}, CMBR (cosmic microwave background radiations) \cite{ref4} 
anisotropy and Plank collaboration \cite{ref5} indicated that our universe is expanding at faster rate. 
Mysterious dark energy (DE) \cite{ref6,ref7} is one of the cause behind this accelerated expansion. According 
 the prediction of CMBR, the universe is having flat geometrical shape on a large scale\cite{ref8}. 
DE(dark energy) may be the cause of this flatness due to a lack of matter in the cosmos \cite{ref9,ref10,ref11,ref12}.
Many researchers have proposed various hypotheses to explain the occurrence of dark energy \cite{ref13,ref14,ref15,ref16,ref17}.
The Cosmological Constant ($\Lambda$) is thought to be the most effective DE option for describing the universe's expansion.
 Various hypotheses have been developed to explain the nature of the cosmological constant \cite{ref19,ref20,ref21,ref22,ref23}. \\

Holographic dark energy (HDE) gets much attention from researchers from many proposed DE models. This is due to its direct relationship 
with space-time. This holographic dark energy supports in explanation of cosmic features of vacuum energy. It is normally accepted that 
the HDE models with the Hubble radius as the IR (Infra-red) cut-off do not follow the current cosmological accelerated expansion 
in FRW metric, while for the event horizon as the IR cut-off, cosmic acceleration of the universe exists \cite{ref24}. Three popular HDE 
models, namely the Ricci scale, future event horizon, and Granda-Oliveros (GO) IR cut-offs, are investigated by Akhlaghi \cite{ref25} 
showing the accelerated expansion and growth of the universe. In this direction, Ghaffari \cite{ref26} examined the cosmological models of HDE 
with GO cut-off. In black hole thermodynamics, when the vacuum energy of a black hole is not greater than its mass, then Horizon length $L$ 
is chosen as the IR cutoff. And the acquired vacuum energy is identified as HDE \cite{ref27}. The holographic principle states that degrees of 
freedom are determined by bounding area rather than volume. \\

It should be noted that the entropy-area relationship of black holes affects the definition and derivation of holographic energy density 
($\rho_D = 3 c^2 \frac{M^{2}_p}{L^2} $). Following quantum effects influenced by Loop Quantum Gravity (LQG), the definition of the 
entropy-area relationship can be changed. When dealing with the entanglement of quantum fields in and out of the horizon, the power-law correction 
is a fascinating modification (correction) of the entropy-area S(A) relation \cite{ref28,ref29}.

The particular form of the power-law corrected entropy-area relation $ S(A)$ is given as follows:
\begin{equation}
	\label{1}
	S(A) = c_0 \left(\frac{A}{a^{2}_1} \right) \left[1+ c_{1} f(A)  \right]
\end{equation}

Here, f(A) is considered by the power-law relation as
\begin{equation}
	\label{2}
	f(A) = \left(\frac{A}{a^{2}_1} \right)^{-\gamma}
\end{equation}
where $c_0$ and $c_1$ are constants, $a_1$ is the UV cut-off at the horizon, and $\gamma$ is a fractional power dependent on the degree of ground 
and excited state mixing. The contribution of the term $f(A)$ to the entropy $S(A)$ may be considered basically minor over a large horizon area 
(i.e., for $ A \gg {a^{2}_1}$ ), and the mixed state entanglement entropy asymptotically approaches the ground state (Bekenstein-Hawking) entropy.
The following relation may be used to produce the formulation of the entropy area relation S(A) for power-law corrected entropy: 


\begin{equation}
	\label{3}
	S(A)=\frac{A}{4 G}\left(1-K_{\alpha} A^{1-\alpha / 2}\right)
\end{equation}
with $\alpha$ is a dimensionless constant parameter, the constant $K_\alpha$ is considered as
\begin{equation}
	\label{4}
	K_{\alpha}=\frac{\alpha(4 \pi)^{\alpha / 2-1}}{(4-\alpha) r_{c}^{2-\alpha}}
\end{equation}

The cross-over scale is indicated by the phrase $r_{c}$. Furthermore, we know that the quantity $A = 4 \pi R_{h}^{2}$  suggest the horizon's 
area (with the horizon radius $R_{h}$). The second term in Eq. (3) gives the power-law correction to the entropy-area law.  The entropy 
remains a well-defined quantity for positive value of parameter $\alpha$ i.e. ($\alpha>0$). Inspired by the relation given in eq. (3), a new 
form of HDE (called the Power-Law Entropy-Corrected HDE (PLECHDE)  \cite{ref35,ref35a}) model was recently presented as follows:
\begin{equation}
	\label{5}
	\rho_D=3 n^{2} M_{p}^{2} L^{-2}-\beta M_{p}^{2} L^{-\delta}
\end{equation}

$\delta$ is a positive exponent, and $\beta$ is a dimensionless parameter. The preceding equation provides the well-known holographic
energy density in the exceptional case $\beta = 0$. The value of $\delta$ determines the significance of the corrected term in 
distinct locations. When $\delta = 2$, the two terms can be merged, recovering the standard HDE density. Let's look at the $\delta > 2 $ and $\delta<2 $
cases independently. In the first situation, for $\delta > 2$, the corrected term can be comparable to the first term only when $L$ is very small.
The range of $\delta$ was claimed as $2 < \delta < 4$ \cite{ref28}. On satisfaction of the generalized second law of thermodynamics for the universe
with the power-law corrected entropy, the second situation $\delta < 2 $ has been rejected \cite{ref29}. Further, Sheykhi et al.\cite{ref35} 
describe the Power-law entropy corrected holographic dark energy model for interacting and non-interacting scenarios. \\

Here, $\beta$ is a dimensionless constant whose precise value need to be determined. Various authors have considered distinct value of 
the parameter $\beta$ suitable for their cosmological models. Karami et al.\cite{ref35b} consider the value $\beta = -14.8$ to describe dynamics of 
agegraphic dark energy model. Karami and Abdolmaleki \cite{ref35c} have discuss the agegraphic dark energy model in $f(T)$ modified teleparallel 
gravity by assuming the values $\beta = 0.1, \beta = -10$ and $\beta = -14.8$. Recently, Jawad et al. \cite{ref35d} developed the Entropy corrected 
holographic dark energy models in modified gravity model utilizing $\beta = 0.5$. Following above mentioned analysis, we have described the dynamics 
of our model using the value $\beta = 0.001$. \\

The basic behaviour of the PLECHDE model with observational confrontation is investigated in this work.
The authors discuss 28 observational Hubble observations in the region of redshift  $0.07 \leq z \leq 2.3 $, see in the reference \cite{ref30}.
For dark energy restrictions, Amirhashchi and Yadav \cite{ref31} listed 31 H(z) observations.
To estimate the model parameter, Amirhashchi et al. \cite{ref32} used 36 H(z) data paired with JLA (Joint Light-curve Analysis) data.
Goswami et al. \cite{ref33} employed observational data from 38 H(z) and 581 SN Ia.
We employed 46 observational Hubble data points in the red-shift region $0 \leq z \leq 2.36$ in this work \cite{ref34,ref34a,ref34b,ref34c}.
\\


The objective of this described model is to consider Power-law entropy corrected Holographic dark energy by assuming IR cut-offs.
Many cosmologists have proposed different parametrization of cosmological parameters, where the model parameters involved in the 
parametrization can be constrained by observational data, in order to describe certain phenomena of the universe, such as phase 
transition from early inflation era to late acceleration phase.
We examine the Hubble parameter parametrization in the manner \cite{ref34d,ref34e,ref34f}: 
$H(a) = \alpha_{1}(1 + a^{-n})$; where $\alpha_{1} > 0$ and $n>0$ are constants. On integration of this parametrization, ewe get an explicit
form of the scale factor as $a(t) = \left(c_{1}e^{n\alpha_{1}t} -1\right)^{1/n}$; where $c_{1} > 0 $ is the constant of integration. 
The present explicit form of scale factor is an exponential function containing two model parameters $n$ and $\alpha_{1}$ which describe the dynamics 
of the universe. As $t \to 0$, we can have $a(0)= (c_{1} - 1)^{1/n}$, which provides a non-zero initial value of scale factor for $c_{1} \ne 1 $ 
(or a cold initiation of universe with finite volume).
The manuscript is structured as: Section $2$ presents the field equations of PLECHDE as IR cut-off. In Section $3$, 
we proposed the solution of field equations. In Section $4$, we describe an observational confrontation with recent H(z) data on the model parameter.
In subsections $5.1$ and $5.2$, we explain the behavior of deceleration parameter $q$ versus redshift $z$ and Statefinder, respectively. Finally, the 
conclusions are summarized in section $6$.

\section{Power law entropy corrected Holographic dark energy as IR cut-off  }

Homogeneous and anisotropic FRW metric form as
\begin{equation}
	\label{6}
	d s^{2}=d t^{2}-a^{2} (t) [d x^{2} + d y^{2} + d z^{2}]
\end{equation}
where a(t) is scale factor and function of $t$.\\

The Einstein’s field equations (EFEs) in General Relativity (GR) can be written as:

\begin{equation}
	\label{7}
	R^{i}_{j}-{\frac{1}{2}}Rg_{ij} = 
	-\frac{8 \pi G}{c^4}( T_{ij} + \bar{T_{ij}} ),
\end{equation}

The energy momentum tensor for PLECHDE and matter for physical interpretation can redefined as: 
$\bar{T_{ij}} =\left({\rho_{{\Lambda}}} + {p_{{\Lambda}}} \right) u_{i} u_{j} + g_{i j} p_{{\Lambda}} {T_{ij}} = \rho_{m} u_{i} u_{j} $, 
where $\rho_{\Lambda}$ and $\rho_{m}$ represents PLECHDE and matter energy densities respectively, and $p_{{\Lambda}}$ is the PLECHDE pressure.\\

The field equations for the discussed metric can be written as:
\begin{equation}
\label{8}
	3 \left(\frac{\dot a}{a}\right)^2 = \rho_{m} + \rho_{\Lambda}
\end{equation}

\begin{equation}
	\label{9}
	\left( 2 \frac{\ddot a}{a} + \left(\frac{\dot a}{a}\right)^2 \right) = - p_{\Lambda},
\end{equation}
The constants for the Einstein equations are assumed as $8 \pi G = 1 = c$. \\

Taking IR cut-off as Hubble horizon, the power-law entropy corrected holographic dark energy (PLECHDE) is defined as:
\begin{equation}
	\label{10}
	\rho_\Lambda = 3 c^2 {M^{2}_p} H^{2} - \beta {M^{2}_p} H^{\delta} ,
\end{equation}
where $M^2_{p}$ = $\frac{1}{8 \pi G}$ is the reduced plank constant and  $3 c^2$ is a numerical constant in the above relation. 
For $\delta > 2$, it was shown that the generalized second law of thermodynamics is conformed for the universe with power-law corrected entropy. 
As a result, the correction has a physical meaning only in the early cosmos, and it becomes meaningless as the universe grows larger \cite{ref35}.\\

\section{Solutions of the field equations for PLECHDE  }
Type-Ia supernova observations \cite{ref2,ref3}, WMAP (Wilkinson Microwave Anisotropic Probe) Collaboration \cite{ref4} and Planck Collaboration \cite{ref5}, have been 	discussed about time-dependent DP which shows decelerating expansion in past, and accelerated expansion at present i.e. there is transition 
from decelerating to accelerating phase. So, to proposed an explicit solution for model of Universe, we have assumed a time dependent scale 
factor in the from \\

\begin{equation}
	\label{11}
	a(t) = (c_{1}e^{n \alpha_1 t} - 1)^{\frac{1}{n}},
\end{equation}
where $n$, $\alpha_1$ and $c_{1}$ are positive constants i.e. ( $n>0$, $\alpha_1 > 0$, $c_{1} \ne 1$).\\

This scale factor suggest a model of transiting universe. The Hubble's and deceleration parameters are determined as 
\begin{equation}
	\label{12}
	H  = \frac{\dot{a}}{a} = \alpha_{1} \left( \frac{c_{1}e^{n \alpha_1 t}}{c_{1}e^{n \alpha_1 t } -1 } \right)
\end{equation}
\begin{equation}
	\label{13}
	q = -\frac{a\ddot{a}}{\dot{a}^2}=-\left( 1+\frac{\dot{H}}{H^2}\right) = -1 + \frac{n}{c_{1}e^{n \alpha_1 t}}
\end{equation}
Eq. (13) shows that the DP is time-dependent, which can take both positive and negative values representing early decelerating phase and 
later accelerating phase. From Eq. (13) we can see that $q = -1 + \frac{n}{c_{1}}$ as $t \to 0$, which is constant and positive for 
$n>1$ and $c_{1} < n$ and for $c_{1} > n$, DP possesses negative value. This indicates that DP has a signature flipping nature from 
positive to negative era with the evaluation.\\

The scale factor and redshift $z$ are linked by the relationship $ a = \frac{a_0}{1+z}$. In terms of redshift $z$, we obtain 
the Hubble parameter defines as 

\begin{equation}
	\label{14}
	H = \frac{H_0}{2} [1 + (1+z)^n],
\end{equation}
Deceleration parameter is defined as
\begin{equation}
	\label{15}
	q= -\frac{\ddot a}{a H^2}  = -1 + \frac{(1+z)}{H(z)} \frac{d H(z)}{dz} ,
\end{equation}
which is equivalent to

\begin{equation}
	\label{16}
	q=  -1 - \frac{n (1+z)^{(n-1)}}{\left[1 + (1+z)^n\right]^2} ,
\end{equation}

From Eqs.(\ref{10}) and (\ref{16}), energy density of PLECHDE in term of redshift $z$ 

\begin{equation}
\label{17}
\rho_\Lambda = 3 c^2 {M^{2}_p} \left[ \frac{H_0}{2} [1 + (1+z)^n] \right]^{2} - \beta {M^{2}_p} \left[ \frac{H_0}{2} [1 + (1+z)^n] \right]^{\delta} ,
\end{equation}

We get matter energy density vs redshift $z$ from Eqs. (\ref{8}) and (\ref{16})

\[\rho_m = 3 \left[\frac{H_0}{2} \left( 1 + (1+z)^n \right)   \right]^2 - \rho_\Lambda
\]
\begin{equation}
	\label{18} 
	= (1 + 3 c^2 {M^{2}_P}) \left[ \frac{H_0}{2} [1 + (1+z)^n] \right]^{2} - \beta {M^{2}_P} \left[ \frac{H_0}{2} [1 + (1+z)^n]  \right]^{\delta}.
\end{equation}

The conservation for PLECHDE is expressed as $\dot \rho_{\Lambda} + 3H ( \rho_{\Lambda} + p_{\Lambda} ) = 0$. The equation of state 
(barotropic) is $p_\Lambda = \omega_{\Lambda} \rho_{\Lambda}$. Using Eq. (\ref{8}), we find $\omega_{\Lambda}$ as :

\begin{equation}
\label{19}
\omega_{\Lambda} = -1 - \frac{\dot H}{3 H^2} \frac{\left( 6 c^2 H^2 - \delta \beta H^{\delta}\right)}{\left( 3 c^2 H^2 - \beta H^{\delta}\right)}
	=  -1 - \frac{2 n}{3 H_0} \frac{(1+z)^{n-1}}{(1 + (1+z)^n)^2}
\left[ \frac{6 c^2 - \delta \beta  \left(\frac{H_0}{2} [1 + (1+z)^n]  \right)^{\delta -2}} {3 c^2  - \beta  \left(\frac{H_0}{2} 
[1 + (1+z)^n] \right)^{\delta - 2} } \right] . 
\end{equation}

\section{Observational constraints on the model parameters}

The observational data and statistical methodological analyses used to restrict the model parameters of the generated universe are presented 
in this section.\\

{\bf Hubble parameter H(z)} \\

With the help of Hubble Space Telescope (HST) \cite{ref36}, Cepheid variable observations \cite{ref37}, gravitational lensing \cite{ref38}, and  
WMAP seven year data \cite{ref39}, many astrophysical researchers \cite{ref40} estimated Hubble constant 
$72 \pm 8$, $69.7_{-5.0}^{+4.9}, 71 \pm 2.5,70.4_{-1.4}^{+1.3}, 73.8 \pm 2.4$ and $67 \pm 3.2$. For more information, see Kumar \cite{ref41}, 
Sharma et al. \cite{ref42}, and Amirhashchi and Yadav \cite{ref31}. For different redshifts, we consider an observed data set of 46 Hubble 
parameters $H_{ob}$ with standard deviations  $\sigma_i$. \\

In this section, we find constraints on the model parameter $ n $ by bounding the model under consideration with recent 46 points of $ H(z) $ 
data set ( OHD) in the redshift range $ 0 \leq z \leq 2.36 $ with their corresponding standard deviation $\sigma_{i} $ and compare with the 
$\Lambda $CDM model. The mean value of the model parameter n determined by minimizing the corresponding chi-square value, which is equivalent 
to the maximum likelihood analysis is given by

\begin{equation}
	\label{20}
	\chi^{2}_{OHD}\left( n\right)=\sum_{i=1}^{46} {\frac{\left(H_{th}(n,z_{i})-H_{ob}(z_i)\right)^2}{\sigma(i)^2}}.
\end{equation}
Here, $ H_{th} $ and $ H_{obs} $ represent the theoretical and observed value of Hubble parameter $ H $ of our model and $n$ is the model parameter.
\begin{table}[H]
	\caption{\small ``The behaviour of Hubble parameter $H(z)$ with redshift}
	\begin{center}
	\begin{tabular}{|c|c|c|c|c|c|c|c|c|c|}
	\hline
	\tiny	$S.No$  &	\tiny  $Z$ & \tiny $H (Obs)$ & \tiny $\sigma_{i}$ & \tiny References & \tiny $S.No$  &	\tiny  $Z$ & \tiny $H (Obs)$ & \tiny $\sigma_{i}$ & \tiny References \\
	\hline
	\tiny	1	& \tiny 0	  &\tiny 67.77 & \tiny 1.30 & \tiny\cite{ref43} & \tiny	24	& \tiny 0.4783	 &\tiny 80.9  & \tiny9 & \tiny \cite{ref51}   \\
			
			\tiny	2	& \tiny 0.07  &\tiny 69    & \tiny 19.6 & \tiny \cite{ref44} & \tiny	25	& \tiny 0.48	 &\tiny 97 & \tiny60 & \tiny \cite{ref46}  \\

			\tiny	3	& \tiny 0.09	 &\tiny 69  & \tiny 12 & \tiny \cite{ref45}   & 	\tiny	26	& \tiny 0.51	 &\tiny 90.4  & \tiny1.9 & \tiny\cite{ref50} \\  
			\tiny	4	& \tiny 0.01	 &\tiny 69  & \tiny 12 & \tiny \cite{ref46}   & \tiny	27	& \tiny 0.57	 &\tiny 96.8  & \tiny3.4& \tiny \cite{ref54}  \\	
			\tiny	5	& \tiny 0.12	 &\tiny 68.6  & \tiny26.2 & \tiny \cite{ref44}  & \tiny	28	& \tiny 0.593	 &\tiny 104 & \tiny 13 & \tiny \cite{ref47}  \\ 
			%
			\tiny	6	& \tiny 0.17	 &\tiny 83  & \tiny 8 & \tiny \cite{ref46}   & \tiny	29	& \tiny 0.60	 &\tiny 87.9  & \tiny6.1 & \tiny\cite{ref52} \\ 
			%
			\tiny	7	& \tiny 0.179	 &\tiny 75  & \tiny 4  & \tiny \cite{ref47}  & \tiny	30	& \tiny 0.61	 &\tiny 97.3  & \tiny2.1 & \tiny\cite{ref50} \\ 		
			\tiny	8	& \tiny 0.1993	 &\tiny 75  & \tiny 5  & \tiny\cite{ref47}   & \tiny	31	& \tiny 0.68	 &\tiny 92  & \tiny 8 & \tiny\cite{ref47}   \\ 
			\tiny	9	& \tiny 0.2	 &\tiny 72.9  & \tiny 29.6  & \tiny \cite{ref44} & 	\tiny	32	& \tiny 0.73	 &\tiny 97.3 & \tiny 7 & \tiny \cite{ref52}   \\ 
			\tiny	10	& \tiny 0.24	 &\tiny 79.7  & \tiny 2.7 & \tiny \cite{ref48} & \tiny	33	& \tiny 0.781	 &\tiny 105 & \tiny 12 & \tiny \cite{ref47}   \\		
			\tiny	11	& \tiny 0.27	 &\tiny 77  & \tiny 14 & \tiny \cite{ref46}   & \tiny	34	& \tiny 0.875	 &\tiny 125  & \tiny 17 & \tiny \cite{ref47}  \\ 
			\tiny	12	& \tiny 0.28	 &\tiny 88.8  & \tiny 36.6 & \tiny \cite{ref44}  & 	\tiny	35	& \tiny 0.88	 &\tiny 90  & \tiny 40 & \tiny  \cite{ref46} \\
			%
			\tiny	13	& \tiny 0.35	 &\tiny 82.7 & \tiny 8.4 & \tiny \cite{ref49}    & \tiny	36	& \tiny 0.9	 &\tiny 117  & \tiny 23 & \tiny  \cite{ref46} \\ 
			\tiny	14	& \tiny 0.352	 &\tiny 83 & \tiny 14 & \tiny \cite{ref47}    & \tiny	37	& \tiny 1.037	 &\tiny 154  & \tiny 20 & \tiny \cite{ref47} \\ 
			\tiny	15	& \tiny 0.38	 &\tiny 81.5  & \tiny 1.9 & \tiny \cite{ref50}  & 	\tiny	38 	& \tiny 1.3	 &\tiny 168 & \tiny 17 & \tiny \cite{ref46} \\ 
			%
			\tiny	16	& \tiny 0.3802	 &\tiny 83 & \tiny 13.5 & \tiny\cite{ref51}  & 	\tiny	39	& \tiny 1.363	 &\tiny 160  & \tiny 33.6 & \tiny \cite{ref55} \\ 
			%
			\tiny	17	& \tiny 0.4	 &\tiny 95  & \tiny 17 & \tiny \cite{ref45}   & \tiny	40	& \tiny 1.43	 &\tiny 177  & \tiny 18 & \tiny  \cite{ref46} \\ 
			%
			\tiny	18	& \tiny 0.4004	 &\tiny 77 & \tiny10.2 & \tiny\cite{ref51}   & 	\tiny	41	& \tiny 1.53	 &\tiny 140  & \tiny  14 & \tiny \cite{ref46} \\ 
			\tiny	19	& \tiny 0.4247	 &\tiny 87.1  & \tiny 11.2 & \tiny\cite{ref51}  & 	\tiny	42	& \tiny 1.75	 &\tiny 202  & \tiny 40 & \tiny \cite{ref46}\\ 
			\tiny	20	& \tiny 0.43	 &\tiny 86.5  & \tiny 3.7 & \tiny\cite{ref46}  & \tiny	43	& \tiny 1.965	 &\tiny 186.5  & \tiny 50.4 & \tiny \cite{ref55}  \\ 
			\tiny	21	& \tiny 0.44	 &\tiny 82.6  & \tiny 7.8 & \tiny\cite{ref50}  & 	\tiny	44	& \tiny 2.3	 &\tiny 224 & \tiny 8 & \tiny\cite{ref56}  \\ 
			\tiny	22	& \tiny 0.44497	 &\tiny 92.8  & \tiny 12.9 & \tiny\cite{ref49}  & 	\tiny	45	& \tiny 2.34	 &\tiny 222 & \tiny 7 & \tiny\cite{ref57} \\
			%
                       \tiny	23	& \tiny 0.47	 &\tiny 89 & \tiny49.6  & \tiny\cite{ref51}    & 	\tiny	46	& \tiny 2.36	 &\tiny 226 & \tiny 8 & \tiny\cite{ref58}'' \\
			\hline	
		\end{tabular}
	\end{center}
\end{table}

\begin{figure}[H]
	\centering
	\includegraphics[width=8cm,height=6cm,angle=0]{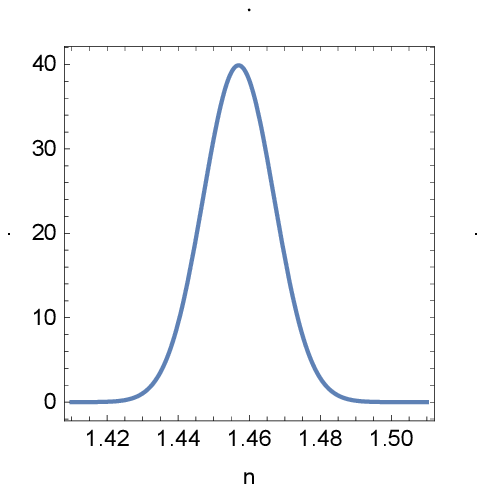}\\
	\includegraphics[width=8cm,height=6cm,angle=0]{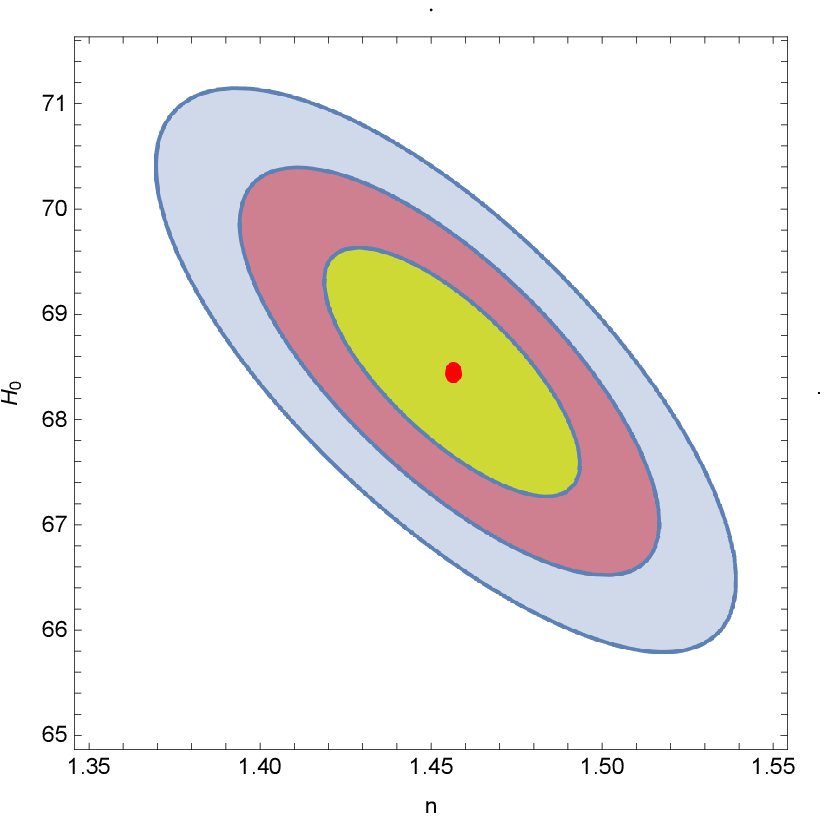}
	\includegraphics[width=8cm,height=6cm,angle=0]{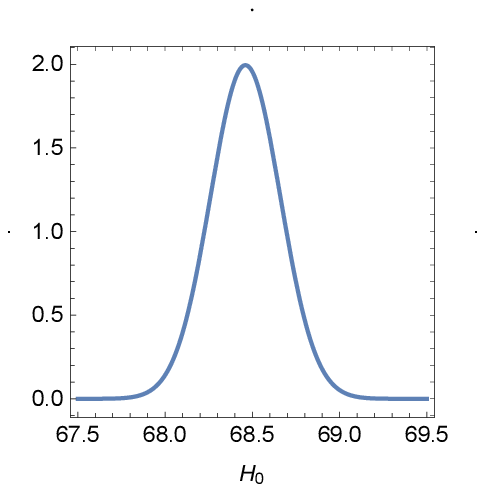}
	\caption{1-Dimensional marginalized distribution and 2-Dimensional contour plots with best fitted values as $n = 1.457\pm0.037$ 
	and $H_0 = 68.53\pm 1.2$ in the $n - H$ plane.}
	\label{fig1}
\end{figure}
Fig.1 shows the two dimensional contours  with 68.3 \%,  95.4 \%  and 99.7 \% confidence level (CL) in $n-H$ plane. From figure, the best fit 
value of parameters are found as $n = 1.457 \pm 0.037$ and $H_0 = 68.53  \pm 1.2 $ .

\begin{figure}[H]
	\centering
	\includegraphics[width=8cm,height=6cm,angle=0]{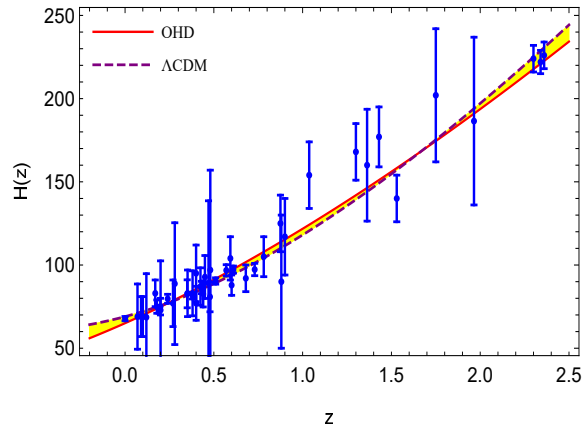}
	~~\caption{The behaviour of our model and the $ \Lambda$CDM model with error bar plots of the Hubble data set.}
	\label{fig2}
\end{figure}

Fig. 2 depicts the comparison of the best-fitting cosmological model of
OHD data and $ \Lambda$CDM with an error bar of Hubble data. We observe that graph for $\Lambda$CDM raise to a quite better fit. 
$H$ increases with the increase of redshift $z$. Here dots signs are 46 observed values of the Hubble constant $(H_{ob})$. The best fit curve 
of the resulting model is represented by the solid red line, whereas the dashed black line represents the comparable $\Lambda$CDM model. 
It's worth noting that the resulting model's predicted value of $H_0$ closely matches the result of the Plank's collaboration \cite{ref59}. 


\section{Physical Behavior of Model}
Here, we have discussed the physical properties of the cosmological model.

\begin{figure}[H]
	(a)\includegraphics[width=7.0cm,height=4.0cm,angle=0]{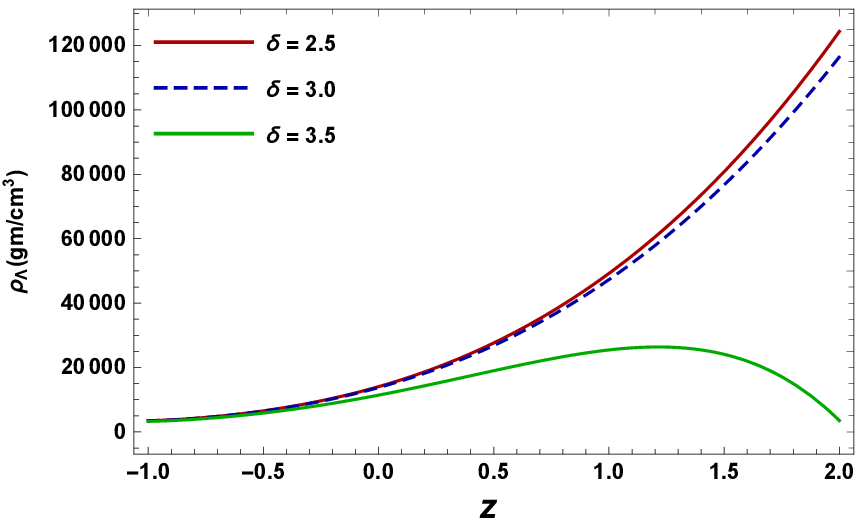}
	(b)\includegraphics[width=7.0cm,height=4.0cm,angle=0]{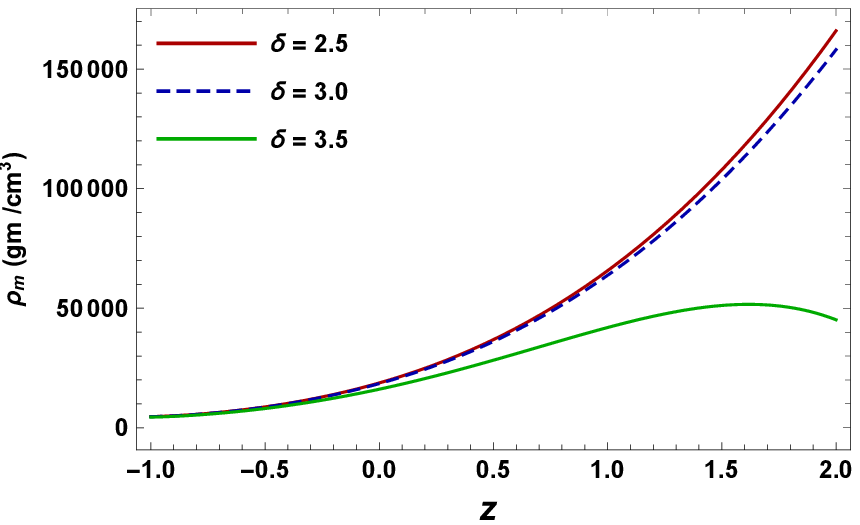}
	\centering
	(c)\includegraphics[width=7.0cm,height=4.0cm,angle=0]{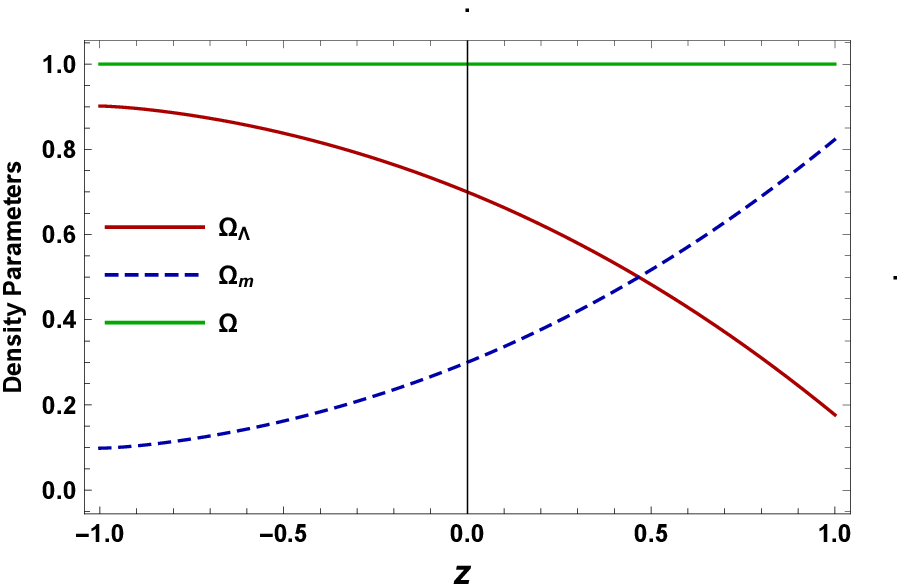}
	\caption{(a) Dark energy density vs redshift $z$ with $\beta = 0.001$, $H_{0} = 68.53$, $n = 1.457$, (b) Matter energy density vs redshift 
	$z$ with $\beta = 0.001$, $H_{0} = 68.53$, $n = 1.457$, 
			(c) Density parameters vs redshift $z$ with $\beta = 0.001$, $\delta=3.6092$, $H_{0} = 68.53$, $n = 1.457$.}
	\label{fig3}
\end{figure}

Figure $3(a)$ and its corresponding Eq. (\ref{17}) portrays PLECHDE dark energy density ($\rho_\Lambda$) with IR cut-off versus redshift $z$ for 
the observational values. It is found to be an increasing function of redshift $z$. For various estimations of $\delta$, dark energy density 
$\rho_\Lambda$ indicates the positive behavior throughout the evolution of the universe. Therefore we observe that our model is stable with 
recent observations. \\
   
Figure $3(b)$ shows matter energy density ($\rho_m$) verses redshift (z) for the observational values $n = 1.457 \pm 0.037$ and $H_0 = 68.53 \pm 1.2 $. 
The figure shows that matter-energy density $\rho_m$ increases slowly and leads to infinity. it is easy to see that the $\rho_m$ is positive 
throughout the region and increases with redshift $z$ for every different value of $\delta$. This result is consistent with observations. 
Using the values $n = 1.457$, $\beta = 0.001$, and $H_{0} = 68.53$, in Eq. (\ref{17}), the $\delta$ is constrained as $3.6092$ for 
the current observational value $\Omega_{\Lambda} = 0.7$.
Figure $3(c)$ plots the variation of density parameters ($\Omega_{\Lambda}, \Omega_{m}, \Omega $) versus redshift $z$. From this figure, 
we observe that the ordinary matter dominates in the early universe, i.e., $\Omega_{m} > \Omega_{\Lambda}$, and it provides a strong physical background 
for the earlier decelerating phase of the universe. But after a transition time, the density parameter for cosmological constant dominates 
the evolution, i.e., $\Omega_{\Lambda} > \Omega_{m}$ which is probably responsible for the accelerated expansion of the present-day Universe. 
The total density parameter ($\Omega$) approaches $1$ for a sufficiently large time, i.e., at the present epoch.

\begin{figure}[H]
	\centering
	\includegraphics[width=7.0cm,height=4.0cm,angle=0]{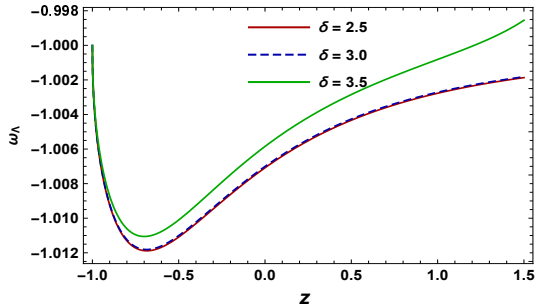}
	\caption{ Plot of $\omega_\Lambda$ vs redshift $z$ with $\beta = 0.001$, $H_{0} = 68.53$, $n = 1.457$. }
	\label{fig4}
\end{figure}
Figure 4 demonstrates the nature of the equation of state parameter $\omega_\Lambda$ concerning redshift $z$ for PLECHDE. it is observed 
that the behavior of the EoS parameter is the same for all values of $\delta$ with the Hubble horizon cut-off. It is rapidly falling at the early stage 
while later on tends to constant value approximate $-1.002$. We have also seen that it lies in phantom region ($\omega_D\le{-1} $) for the 
observational parameters $n = 1.457 \pm 0.037$ and $H_0 = 68.53  \pm 1.2 $. This result is consistent with the recent observational dataset.

\subsection{Deceleration Parameter }

\begin{figure}[H]
	\centering
	\includegraphics[width=6.5cm,height=4.0cm,angle=0]{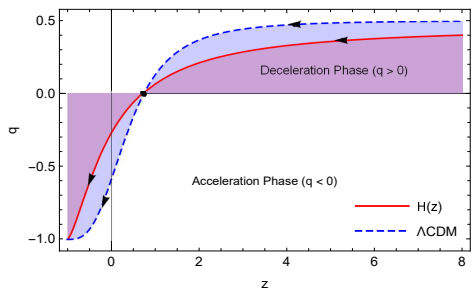}
	~~\caption{Plot of deceleration parameter $q$ versus redshift $ z $ for both H(z) and $\Lambda$CDM models.}
	\label{fig5}
\end{figure}

Fig. 5 depicts the deceleration parameter comparing OHD and $\Lambda$CDM data. The solid red line exhibited the best fit curve of the
derived model, and the dashed blue line displayed the corresponding $\Lambda$CDM model. The universe shows an expansion with the change of signature
flipping from decelerating to accelerating phase at the transition redshift $z_t = 0.71165 $. This transition redshift is well consistent
with recent 36 OHD provided redshift range $0.07 \leq z \leq 2.36$ \cite{ref33}. Comparing 740 SN Ia with JLA indicates a redshift range
$0.01 \leq z \leq 1.30$. Thus our results are in good agreement with recent studies mentioned in Refs. \cite{ref31,ref32,ref33,ref34}.

\subsection{Statefinder diagnosis}

The statefinder pairs $\{r,s\}$ are the geometrical quantities that are directly obtained from metrics. This diagnostic is used to distinguish
different dark energy models and hence becomes an important tool in modern cosmology. Alam et al. \cite{ref61} have defined the statefinder
parameters $r$ and $s$ as following
\begin{equation}
	\label{21}
	r=\frac{\dddot{a}}{a H^3},  \   s=\frac{r-1}{3 (q-\frac{1}{2})}
\end{equation}

A notable feature of the state finder is that these parameters are geometric because they depend on the scale factor and its time
derivative \cite{ref62}. In addition, different dark energy models show different evolutionary trajectories in the $ s-r $ plane.
In addition, the well-known flat $\Lambda CDM$ model corresponds to  points $ s = 0 $ and $ r = 1 $ on the $ s-r $ plane. These properties
of the Statefinder allow you to distinguish between different models of dark energy. In the literature, Statefinder diagnostic tools are
often used to distinguish between different dark energy models \cite{ref63,ref64}.\\

The statefinders can also read as

\begin{equation}
	\label{22}
	r=2 q^{2}+q - \frac{\dot{q}}{H} \quad,  \   s=\frac{2}{3}(q+1) - \frac{\dot{q}}{3H(q-\frac{1}{2})}
\end{equation}
At $q=-1$, we observe as $r=1$, $s=0$ and our cosmic model resembles the $\Lambda$CDM model. For values of $q$ in the range $-1\leq q<0.5$, 
we get the evolutionary $q-r$, $q-s$, and $s-r$ trajectories for the cosmological model, as illustrated in Figs. 6(a), 6(b), and 6(c). 
The flat $\Lambda$CDM model is shown by the black dot in panel (c) at $(s, r)=(0,1)$. The $\Lambda$CDM statefinder pair $(0,1)$ is an attractor 
in our cosmological model, which is interesting to note.


\begin{figure}[H]
	(a)\includegraphics[width=6.0cm,height=4.0cm,angle=0]{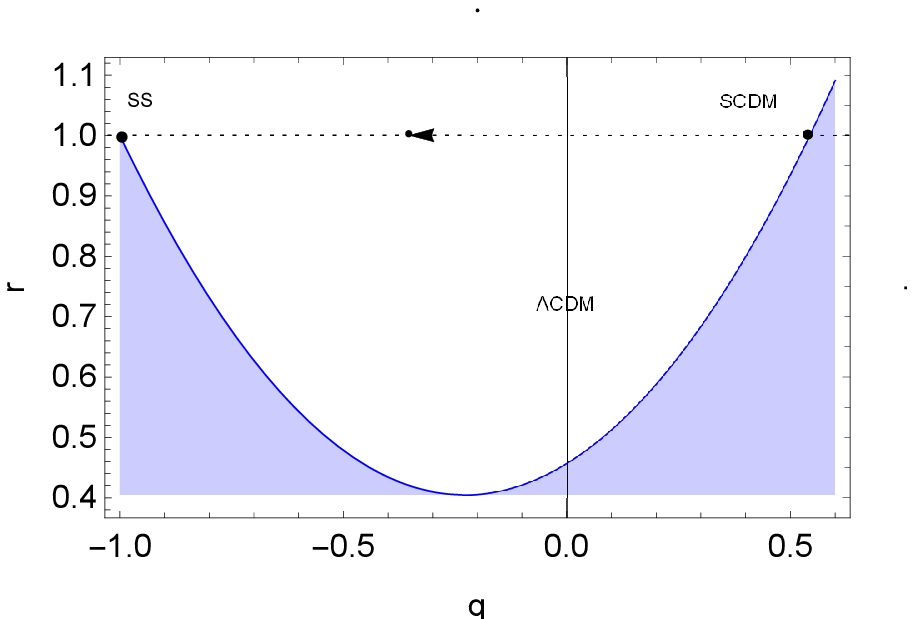}
	(b)\includegraphics[width=6.0cm,height=4.0cm,angle=0]{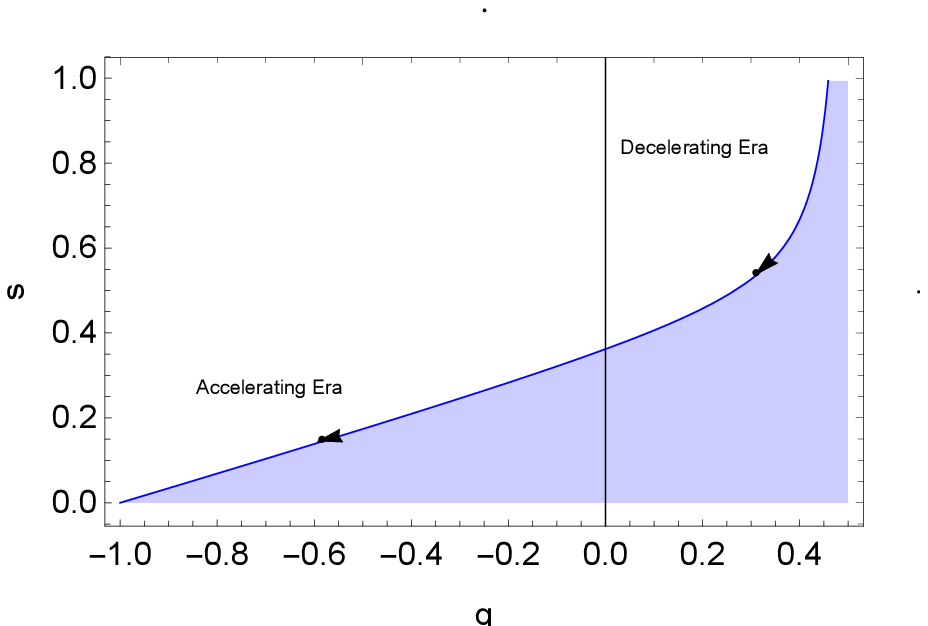}
	\centering
	(c)\includegraphics[width=6.0cm,height=4.0cm,angle=0]{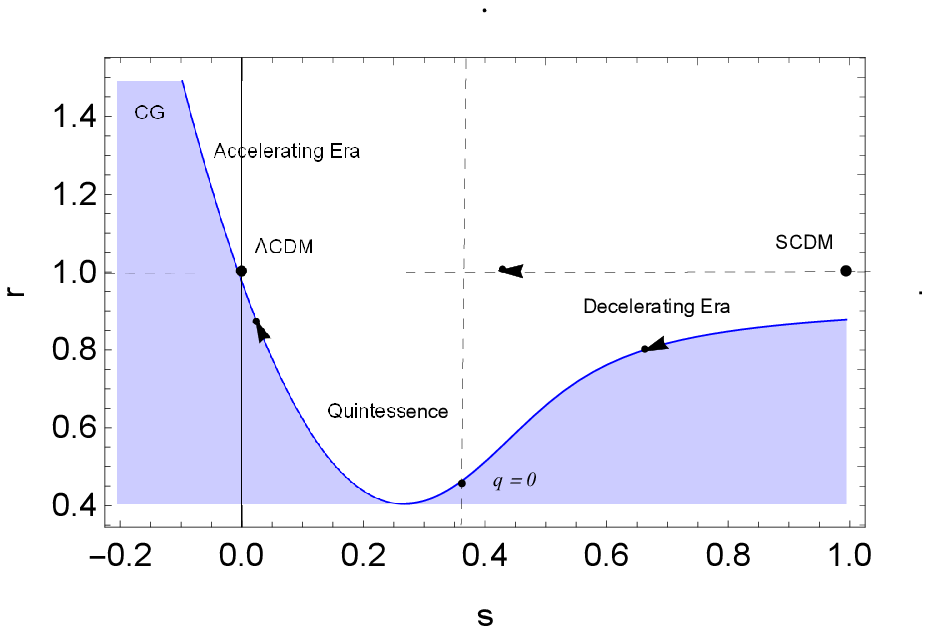}
	\caption{Plots of statefinders in $q-r$ plane (Fig.6a), $q-s$ plane (Fig.6b)  and s-r plane (Fig.6c). Here, black dot represents the location of point $(q, r) = (-1, 1)$ in the $q-r$ plane, $(q, s) = (-1, 0)$ and $(s, r) = (0, 1)$ in the $s-r$ plane. The vertical dashed line separates the acceleration and deceleration zones in the Fig.6a and Fig.6c. The arrows indicate the direction of trajectories' evolution as q varies from $0.5$ to $-1$. }
	\label{fig6}	
\end{figure}

\section{Concluding Summary}
In this paper, we have discussed the transition model of Power-law entropy corrected holographic dark energy in the context of Brans Dicke's theory.
To find the solutions of the field equations, we have precised the energy density of PLECHDE
$\rho_\Lambda = 3 c^2 {M^{2}_p} H^{2} - \beta {M^{2}_p} H^{\delta} $. We have also discussed the evolution of physical and dynamic parameters
of the universe and their cosmological significance. The main  highlights of the model are as follows:

\begin{itemize}
	\item 	
	Figure 1, depicts one-Dimensional marginalized distribution and two-Dimensional contour plots with best fitted values as 
	$n = 1.457\pm0.037$ \& $H = 68.53\pm 1.2$ in the $n - H$ plane.
	
	\item
	Figure 2 shows the comparison of the best-fitting cosmological model of OHD data and $ \Lambda$CDM with the error bar of Hubble data. 
	We observe that graph for $\Lambda$CDM raise to a quite better fit. H increases with the increase of redshift z. Here dots signs 
	are 46 observed values of the Hubble constant $(H_{ob})$. It is worth noting that the derived model estimates $ H_0 $ are in 
	good agreement with the  Plank collaboration results \cite{ref59}.

	\item It has been observed from the figures $3(a)$ and $3(b)$ that for different values of $\delta$, dark energy density 
	$\rho_\Lambda$ and matter energy density $\rho_m$ are increasing function vs redshift $z$ for the observational values 
	$n = 1.457\pm0.037$ \& $H = 68.53\pm 1.2$.
	
	\item Plot 4, explain EoS parameter $\omega_\Lambda$ for the observational values for power law entropy corrected holographic 
	dark energy. It has been plotted for three different values of $\delta$ ($2.00,  2.05, 2.10$). It is found to be negative and 
	lies in phantom region ($\omega_D \leq{-1}  $).
	
	\item 
	Figure 5, explain of deceleration parameter $q$ versus redshift $ z $ for both H(z) and $\Lambda$CDM models. In this derived model, 
	it has been noted that PLECHDE model exhibits a smooth transition from deceleration to current acceleration phase at the transition 
	point $z_t = 0.71165$. The filled circle shows the best fit values of the deceleration parameter at transition redshift. The results obtained 
	in derived are consistent with observational data of modern cosmology, as clearly seen in fig.3.
	\item  
	Figures 6(a), 6(b), and 6(c) show State-finders in the $q-r$ plane, $q-s$ plane, and $s-r$ plane. In the $q–r$ plane, the black dot 
	represents SS and SCD models respectively, and in the $s-r$ plane, the black dot represents $\Lambda$CDM and SCDM models respectively. 
	The vertical dashed line separates the acceleration and deceleration zones in the right and left panels. The arrows indicate the 
	direction of trajectories' evolution as $q$ varies from $0.5$ to $-1$. At $q=-1$, we notice that $r=1$ and $s=0$. As a result, 
	at $q=-1$, our cosmological model resembles the $\Lambda$CDM model, which is consistent with current data.
	
\end{itemize}

Hence, our constructed transit PLECHDE model with observational data has good agreement with recent observations.

\section*{Acknowledgement}
 A. Pradhan thanks the IUCAA, Pune, India, for providing facilities under the associateship program.
Sincere thanks are due to anonymous reviewers for their constructive comments to enhance the quality of the paper.

\section*{Appendix}
From Eqs. (\ref{12}) and (\ref{13}), we obtain
\begin{equation}
	\label{23}
	H = \frac{\dot{a}}{a}
\end{equation}

\begin{equation}
	\label{24}
	q = -\frac{a\ddot{a}}{\dot{a}^2}=-\frac{\dot{H}}{H^2} - 1
\end{equation}
The statefinders are defined as
\begin{equation}
	\label{25}
	r=\frac{\dddot{a}}{a H^3}= \frac{\ddot{H}}{H^{3}} + 3\frac{\dot{H}}{H^{2}} + 1
\end{equation}
Now from (\ref{24})
\begin{equation}
	\label{26}
	\frac{\ddot{H}}{H^{3}} = -\frac{\dot{q}}{H} + 2\frac{\dot{H}^{2}}{H^{4}}
\end{equation}
Putting the values of (\ref{24}) and (\ref{26}) in Eq. (\ref{25}), we get
\begin{equation}
	\label{27}
 r = -\frac{\dot{q}}{H} + 2\frac{\dot{H}^{2}}{H^{4}} - 3(q + 1) + 1
\end{equation}
which reduces to
\begin{equation}
	\label{28}
r = q(1 + 2q)-\frac{\dot{q}}{H}	
\end{equation}
From Eqs. (\ref{21}) and (\ref{28}), we obtain
\begin{equation}
	\label{29}
	s = \frac{r - 1}{3(q -\frac{1}{2})} = \frac{2}{3}(q + 1) - \frac{\dot{q}}{3H(q -\frac{1}{2})}
\end{equation}


\end{document}